\documentclass[amstex,showpacs,showkeys,floats,floatfix,superscriptaddress,aps,pra]{revtex4}
\usepackage{amssymb}
\usepackage{amsmath}
\usepackage{calc}
\usepackage{graphicx}
\usepackage{bm}

\def\be{\begin{equation}}
\def\ee{\end{equation}}

\def\half{\frac{_1}{^2}}

\def\A{\mathsf{A}}
\def\B{\mathsf{B}}
\def\D{\mathsf{D}}
\def\H{\mathsf{H}}

\def\M{\mathsf{M}}

\def\U{\mathsf{U}}
\def\u{\mathsf{u}}
\def\V{\mathsf{V}}
\def\I{\mathsf{I}}

\def\Na{N}
\def\Nb{M}
\def\Nd{N_0}
\def\CKa{a}
\def\CKb{b}
\def\MSa{\alpha}
\def\MSb{\beta}
\def\MSd{\gamma}
\def\EV{\lambda}
\def\ket#1{\vert #1 \rangle}
\def\psia{\psi}
\def\psib{\varphi}

\begin{document}
\author{E. S. Kyoseva}
\affiliation{Department of Physics, Sofia University, James Bourchier 5 blvd., 1164 Sofia, Bulgaria}
\author{N. V. Vitanov}
\affiliation{Department of Physics, Sofia University, James Bourchier 5 blvd., 1164 Sofia, Bulgaria}
\affiliation{Institute of Solid State Physics, Bulgarian Academy of Sciences, Tsarigradsko chauss\'{e}e 72, 1784 Sofia, Bulgaria}
\author{B. W. Shore}
\affiliation{Fachbereich Physik, Technische Universit\"at Kaiserslautern, 67653 Kaiserslautern, Germany}
\affiliation{Permanent address: 618 Escondido Circle, Livermore, CA 94550, USA}
\title{Physical realization of coupled Hilbert-space mirrors for quantum-state engineering}
\date{\today }

\begin{abstract}
Manipulation of superpositions of discrete quantum states has a mathematical counterpart in the motion
 of a unit-length statevector in an $N$-dimensional Hilbert space.
Any such statevector motion can be regarded as a succession of two-dimensional rotations. 
But the desired statevector change can also be treated as a succession of reflections, the generalization of Householder transformations.  
In multidimensional Hilbert space such reflection sequences offer more efficient procedures for statevector manipulation than do sequences of rotations. 
We here show how such reflections can be designed for a system with two degenerate levels
 -- a generalization of the traditional two-state atom -- that allows the construction of propagators for angular momentum states.
We use the Morris-Shore transformation to express the propagator in terms of Morris-Shore basis states and Cayley-Klein parameters,
 which allows us to connect properties of laser pulses to Hilbert-space motion.
Under suitable conditions on the couplings and the common detuning,
 the propagators within each set of degenerate states represent products of generalized Householder reflections, with orthogonal vectors.
We propose physical realizations of this novel geometrical object with resonant, near-resonant and far-off-resonant laser pulses.
We give several examples of implementations in real atoms or molecules.
\end{abstract}
\maketitle



\section{Introduction}


Manipulation of discrete quantum states has long held interest, most recently for application to quantum information processing \cite{QI}. 
In the simplest realizations one deals with a nondegenerate two-state system (a \textquotedblleft two-level atom\textquotedblright \cite{AE,Shore})
 and through pulsed resonant coherent excitation produces a specified superposition of the two states, starting from a single state. 
Following the availability of laser light sources, much attention centered on producing complete excitation,
 i.e. complete population transfer from the ground state to an excited state \cite{Shore}. 
More recently interest has shifted to the production of specific superpositions of the two states, and to producing transitions
 between superposition states. 
In the context of quantum information a two-state system serves as a qubit; more general tasks involve multiple quantum states (a qunit),
 possibly driven by pulsed nonresonant light.

The goal of quantum-state manipulation is to realign the statevector $\Psi(t)$ from some given Hilbert-space direction $\Psi (t_i)$
 at an initial time $t_i$ to some prescribed state $\Psi (t_f)$ at a final time $t_f$, each of these statevectors being defined
 by the set of complex-valued probability amplitudes $C_n(t)$ associated with a set of quantum states $\ket{\psia_n}$. 
Expressed in matrix form, the goal is to obtain a propagator matrix $\U(t_{f},t_{i})$ that transforms the vector $\mathbf{C}(t)$ of components $C_{n}(t)$,  
\begin{equation}
\mathbf{C}(t_f)=\U(t_f,t_i)\mathbf{C}(t_i) .
\end{equation}

As noted in the following section, there  exists an extensive literature describing analytic solutions to the two-state system
 for a variety of pulsed-excitation scenarios. 
There exist also numerous analytic solutions to multistate systems.

Here we consider, as a particular generalization of the two-state system, a two-level system involving two sets of degenerate sublevels:
 a less energetic set of $\Na$ states $\ket{\psia_n}\ (n=1,2,\ldots,\Na)$ with common energy $E_{\Na}$ (the \emph{lower} set),
 and a more energetic set of $\Nb$ states $\ket{\psib_m}\ (m=1,2,\ldots,\Nb)$ that share energy $E_{\Nb}$ (the \emph{upper} set). 
Thus we consider a Hilbert space of dimension $\Na + \Nb$. 
For definiteness we here assume that $\Na\geqq \Nb$. 
We assume that interactions produced by laser fields can induce direct transitions between lower sublevels and upper sublevels,
 but not directly within either manifold of states (i.e. we allow electric-dipole transitions only).

Because the statevector has unit length at all times, any such change can be regarded as a rotation,
 and any allowable motion in several dimensions can be decomposed into a succession of two-dimensional rotations,
 analogous to the three Euler angles that define an arbitrary rotation in Euclidean space.
However, length-preserving changes in multi-dimensional space can also be produced by a succession of reflections. 
When expressed in matrix form these are generalizations of the Householder reflections used in matrix calculus \cite{Householder}. 
As has been shown previously, such unitary operations can be implemented very efficiently using coherent excitation techniques \cite{Kyoseva,Ivanov}; 
 as subsequently noted \cite{Ivanov2}, the use of quantum Householder reflections (QHR) permits efficient
quantum-state engineering of transitions between arbitrary superpositions.

The QHR implementation proposed earlier \cite{Kyoseva,Ivanov} requires a particular multistate linkage pattern,
 in which a set of $\Na$ low-lying degenerate states all link, via radiative interaction, with a \emph{single} upper state
 --  a generalization of the tripod linkage termed  an \emph{N-pod}. 
Such a linkage pattern occurs with a lower level having angular momentum $J=1$ (three sublevels)
 excited to an upper level having $J=0$ (a single sublevel) but the pattern is difficult to realize for more than 3 lower states. 
It is therefore desirable to extend the QHR technique to more general linkage patterns.

The present manuscript describes a procedure that allows implementation of the QHR when there are multiple states in the upper set. 
Such an extension makes possible the application of QHR to arbitrary angular momentum states, as occur with free atoms and molecules. 
As we show, the propagator for such situations is not a single QHR, but a product of QHRs, moreover with orthogonal vectors. 
The expression for the resulting propagator has a clear geometric interpretation as the effect of a succession of reflections,
 i.e. \emph{coupled mirrors}.

The key to this extension of the QHR is a transformation of the underlying basis states,
 the so-called Morris-Shore (MS) transformation \cite{MS,Vitanov MW,other MS}. 
This replaces the original system, with its multiple linkages between states, by a set of independent nondegenerate two-state systems, 
 thereby allowing us to utilize the considerable literature of analytic solutions to two-state systems
 and produce analytic solutions for degenerate multistate systems \cite{Kyoseva}.

In this paper we present the solution of the degenerate two-level problem in a simple closed form involving sums of projectors of MS states. 
These expressions are useful for deriving analytical solutions, generalizations of known two-state solutions, for systems having degenerate levels. 
For certain conditions, wherein the transition probabilities between states from different sets vanish,
 the propagator for states of the lower set is given by a product of QHRs with orthogonal vectors, each vector being a bright state;
 a similar property applies to the upper set.
We discuss properties of this novel geometric object, the coupled mirrors.  
To illustrate the procedure we develop a useful explicit analytic formalism for two upper states and present some examples.


\section{The degenerate two-level system}



\subsection{Two non-degenerate states}


We consider the controlled alteration of  a multistate system, expressed as a redirection of the statevector, induced by a set of laser pulses. 
The simplest example of coherent excitation occurs when there are just two nondegenerate states, indexed 1 and 2. 
The time-dependent Schr\"{o}dinger equation prescribes the changes in Hilbert space as 
 (here and henceforth we set $\hbar = 1$, thereby making no distinction between energy and frequency units) 
\begin{equation}
i \frac{d}{dt}\mathbf{C}(t)=\H(t)\mathbf{C}(t),
\end{equation}%
where the column vector $\mathbf{C}(t)$ has the complex-valued probability amplitudes $C_n(t)$ as elements, $\mathbf{C}(t)=[C_1(t),C_2(t)]^T$.
Commonly one neglects in the Hamiltonian $\H(t)$ elements that vary rapidly compared with the characteristic response times of the system,
 the so-called rotating-wave approximation (RWA) \cite{Shore}. 
Then the Hamiltonian matrix, with suitable choice of energy zero-point, reads
\begin{equation}
\H(t)=\left[ \begin{array}{cc}
0 & \half \Omega(t) \\ 
\half \Omega(t)^{\ast } & \Delta (t)%
\end{array}\right] .
\end{equation}%
Here the slowly-varying function  $\Omega(t)$, the Rabi frequency, quantifies the coupling between the two states. 
For a laser-driven electric-dipole transition in an atom or a molecule,
 $\Omega(t)$ is proportional to the transition dipole moment and the electric-field envelope. 
The detuning $\Delta(t)$ measures the frequency offset of the carrier laser frequency $\omega$
 from the Bohr transition frequency $\omega_0$, $\Delta =\omega_0-\omega $.
A time dependence in the detuning can be introduced by both $\omega$ (due to chirping \cite{femto}) and $\omega_0$ (due to Stark and Zeeman shifts).

When the transition is exactly resonant the associated detuning vanishes, $\Delta =0$. 
Then the response depends only upon the initial conditions and upon the time-integrated Rabi frequency (the temporal pulse area), 
\begin{equation}
A(t)= \int_{0}^{t}dt^{\prime } \, |\Omega(t^\prime)|.
\end{equation}
At all times the probability amplitudes are expressible as trigonometric functions of $A(t)$;
 when $A(t)$ is some even-integer multiple of $\pi $ (e.g., a $2\pi $\emph{\ }pulse), the probabilities repeat their initial values. 
Thus the response, for resonant excitation, does not depend on any details of the pulse shape. 
Of particular use are the $\pi $ pulses, which produce complete population inversion between the two states,
 and half-$\pi $ pulses that create an equal coherent superposition of the two states. 
Resonant pulses having precise temporal area have had wide application, most notably in nuclear magnetic resonance \cite{NMR}
 and coherent atomic excitation \cite{Shore}. 
They are now a common tool in quantum information processing \cite{QI}.

A variety of nonresonant ($\Delta\neq 0$) pulses also lead to exact analytic
expressions for the probability amplitudes $C_{1}(t)$ and $C_{2}(t)$.
Amongst the soluble two-state models are the models of Rabi \cite{Rabi},
 Landau-Zener \cite{LZ}, Rosen-Zener \cite{RZ}, Allen-Eberly \cite{AE,Hioe}, Bambini-Berman \cite{BB},
 Demkov-Kunike \cite{DK}, Demkov \cite{Demkov}, Nikitin \cite{Nikitin}, and Carroll-Hioe \cite{CH}. 
Methods for approximate solutions are also available, such as the perturbation theory, the adiabatic approximation \cite{STIRAP},
 the Magnus approximation \cite{Mag54}, and the Dykhne-Davis-Pechukas approximation \cite{DDP}. 
The latter, in particular, is a very useful tool for deriving very accurate approximations
 in various cases of interest, e.g., transform-limited \cite{Vasilev-Gaussian} and chirped Gaussian pulses \cite{Vasilev-Gaussian-chirped}.
The two-state dynamics acquires interesting new features when the pulsed field possesses some symmetries \cite{symmetry},
 or when it is a sequence of identical pulses (pulse train) \cite{train}.

The two-state atom provides a very basic model of coherent excitation, widely used because of the relative simplicity
 with which it can be treated mathematically and fabricated experimentally. 
However, with this mastery has come interest in more general systems, involving more than two discrete quantum states. 
It is with a class of these that the present paper deals. 
Specifically, we describe a straightforward technique that allows efficient redirection of the statevector for such a system
 by means of laser pulses, and the use of the many exact and approximate solutions available for two-state systems.


\subsection{Multiple states as two degenerate levels}


An important extension is the linear chain of $N$ states, each state being linked only to its nearest neighbor. 
The simplest of this is the three-state chain, involving two independent pulses,
 the mathematics of which has been extensively reported \cite{three state}, as have lengthier chains \cite{chains}. 
Here we consider an extension in which each of the states of the two-state system is replaced by a degenerate set of states
 that together form an energy \emph{level}. 
Such situations occur commonly when one deals with angular-momentum eigenstates, as happens with the electronic structure of atoms in free space.
For a given angular momentum $J$ there are $2J+1$ magnetic sublevels that, in the absence of an external electric or magnetic field, all have the same energy.

\begin{figure}[tb]
\includegraphics[width=80mm]{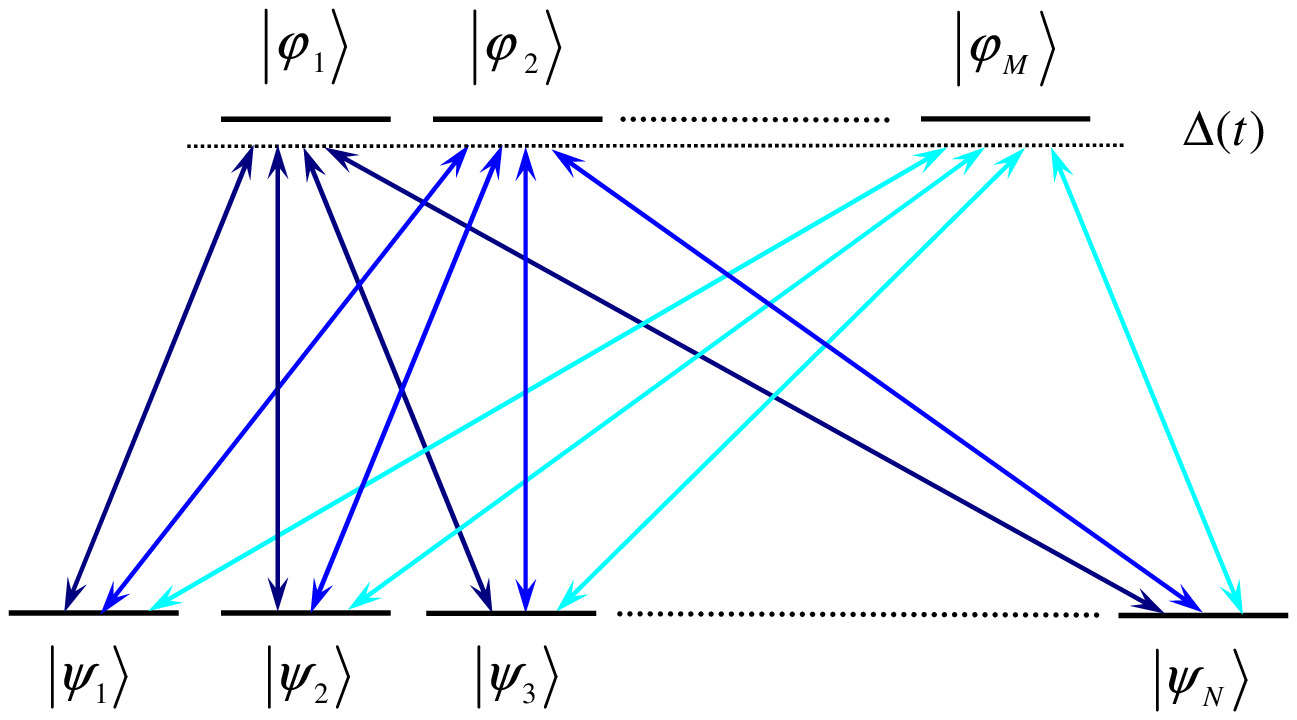}
\caption{(Color online) Schematic linkage pattern for multistate system
consisting of two coupled sets of degenerate levels. 
}
\label{Fig-link}
\end{figure}

We shall consider the possibility that there be no nonzero elements of the Hamiltonian matrix linking any states of the same energy;
 nonzero couplings occur only between states from the lower set and the states in the upper set. 
We make no restriction on the number of states that connect with any single state. 
This generalizes the usual situation of angular momentum states excited by electric-dipole radiation,
 when any state can link to no more than three other states (the selection rules $\Delta m = -1, 0, +1$). 
Figure \ref{Fig-link} illustrates the linkage pattern we consider.

We shall assume that we can control the magnitude and the phase of each interaction (e.g. each electric field envelope),
 and that \emph{every Rabi frequency has the same time dependence}, which we write as 
\begin{equation}
\Omega_{mn}(t)= 2V_{mn} f(t),  \label{Omegas}
\end{equation}
where $V_{mn}$ is a complex-valued constant and $f(t)$ a real-valued pulse-shaped function of time bounded by unity. 
We assume that the duration of the pulse is shorter than any decoherence time, so that the dynamics is governed by the Schr\"{o}dinger equation. 
Such is the situation for excitations of free atoms and molecules by picosecond or femtosecond pulses \cite{femto}.

We shall also assume that each transition has exactly the same detuning $\Delta $. 
In the simplest situations, those associated with angular-momentum states,
 three distinct fields can be distinguished by their polarization, while sharing a common carrier frequency. 
By utilizing three independent polarization directions of a laser pulse we can ensure three linkages with common time dependence. 
More general linkage patterns, still within the model described here, are also possible \cite{other MS}.


\section{Exact analytical solution\label{Sec-exact analytical solution}}



\subsection{The RWA Hamiltonian}


In the RWA the Schr\"{o}dinger equation provides a prescription for the time dependence of the probability amplitudes $C_n(t)$. 
Written in matrix form it is 
\begin{equation}
i \frac{d}{dt}\mathbf{C}(t)=\H(t)\mathbf{C}(t),  \label{SEq}
\end{equation}
 where the column vector $\mathbf{C}(t)$ has $C_n(t)$ as elements. 
We shall arrange these elements with those of the lower set first, followed by the upper set.
Then the RWA Hamiltonian has the block structure 
\begin{equation}
\H(t)=\begin{bmatrix}
\mathsf{O} & \V f(t) \\ \V^\dag f(t) & \D(t)
\end{bmatrix}.  \label{H}
\end{equation}
Here  $\mathsf{O}$ denotes the $\Na$-dimensional square null matrix;
 the zeros signify that the lower states do not interact with each other and that they all have the same energy,
 which we take as the zero-point of our energy scale. 
The matrix $\V $ is an $(\Na\times \Nb)$-dimensional matrix whose elements $V_{nm}$ are the magnitudes of the couplings
 between the lower and the upper states; $\V ^{\dag }$ is its hermitian conjugate. 
Lastly, $\D(t)$ is an $\Nb$-dimensional square diagonal matrix whose elements are all equal to the shared detuning $\Delta (t)$, 
\begin{equation}
\D(t)=\Delta (t)\I,  \label{D}
\end{equation}
where $\I$ is the unit matrix.
The diagonal nature of $\D$ indicates the absence of interaction of the  upper states amongst themselves. 
We allow the detuning $\Delta(t)$ to vary with time, bearing in mind use of known analytic solutions to the two-state model with frequency-swept detuning.
It proves useful to write the matrix of interactions $\V $ as a row vector of $\Na$-dimensional column vectors
 $\left\vert V_{n}\right\rangle =\left[V_{1n},V_{2n},\ldots ,V_{\Na n}\right] ^{T}$,
\begin{equation}
\V  = \begin{bmatrix}
V_{11} & V_{12} & \cdots & V_{1\Nb} \\ 
V_{21} & V_{22} & \cdots & V_{2\Nb} \\ 
\cdots & \cdots & \ddots & \vdots \\ 
V_{\Na1} & V_{\Na2} & \cdots & V_{\Na\Nb}%
\end{bmatrix}
 =[\left\vert V_{1}\right\rangle ,\left\vert V_{2}\right\rangle,\ldots ,\left\vert V_{\Nb}\right\rangle ].  \label{V}
\end{equation}


\subsection{The Morris-Shore transformation}

\begin{figure}[tb]
\includegraphics[width=80mm]{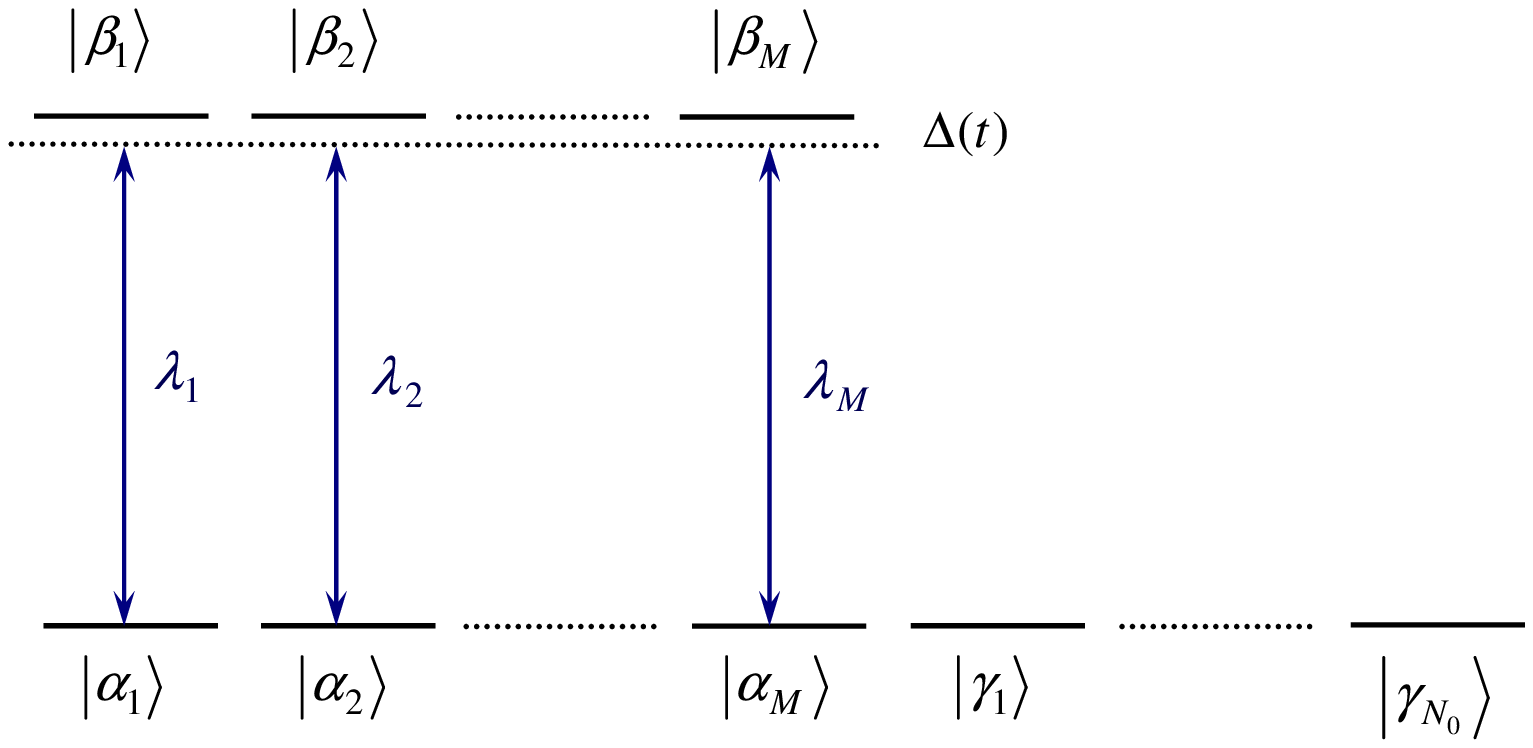}
\caption{(Color online) Linkages of Fig. \protect\ref{Fig-link} transformed by the Morris-Shore transformation into a set of
  $\Nb $ independent nondegenerate two-state systems,  with couplings $\protect\EV_n$,
 and a set of  $\Nd = \Na - \Nb$ decoupled (dark) states. }
\label{Fig-MS}
\end{figure}

A significant property of the linkage pattern shown in Fig. \ref{Fig-link} is that a transformation of Hilbert-space coordinates
 -- the Morris-Shore (MS) transformation -- reduces the dynamics to that of a set of independent
 two-state systems, together with decoupled states \cite{MS}. 
When $\Na>\Nb$, as is the case with Fig. \ref{Fig-link}, the $\Nd=\Na-\Nb$ additional states are a part of the lower-level manifold. 
They have no connection with excited states, and hence they cannot produce excitation followed by fluorescence;
 they are termed \emph{dark states}  \cite{dark state}, by contrast to the states that can produce excitation (and thence fluorescence),
 the \emph{bright states}. 
The number of bright states is the lesser of $\Na$ and $\Nb$, in this case $\Nb$.
Figure \ref{Fig-MS} shows the new linkage pattern in the MS basis.

Because, by assumption, all elements of the interaction matrix share a common time dependence,
 the required MS transformation is achieved by a constant unitary transformation  
\begin{equation}
\mathsf{S}=\left[ \begin{array}{cc}
\A & \mathsf{O} \\ 
\mathsf{O} & \B%
\end{array}\right] .  \label{W}
\end{equation}%
Here $\A$ is a unitary $\Na$-dimensional square matrix ($\mathsf{AA}^{\dagger }=\A^{\dagger }\A=\I$) which transforms the lower set of states,
 and  $\B$ is a unitary $\Nb$-dimensional square matrix ($\B\B^{\dagger }=\B^{\dagger }\B=\I$) which transforms the upper set.
This transformation casts the dynamics into the MS basis,
 with new MS bright lower states $\{|\MSa_m\rangle\}_{m=1}^M$, dark lower states $\{|\MSd_k\rangle\}_{k=1}^{\Nd}$,
 and upper states $\{|\MSb_m\rangle\}_{m=1}^M$. 
The transformed MS Hamiltonian has the form
\begin{equation}
\widetilde{\H}(t)=\mathsf{S}\H(t)\mathsf{S}^{\dagger }=\left[\begin{array}{cc}
\mathsf{O} & \widetilde{\V }f(t) \\ 
\widetilde{\V }^{\dagger }f(t) & \D(t)%
\end{array}\right] ,  \label{H-MS}
\end{equation}%
where%
\begin{equation}
\widetilde{\V }=\A\V\B^{\dagger }.
\label{V-MS}
\end{equation}

Because  the matrices  $\widetilde{\V }\widetilde{\V }^{\dagger }=\A\V \V ^{\dagger }\A^{\dagger }$
 and $\widetilde{\V }^{\dagger }\widetilde{\V }=\B\V ^{\dagger }\V \B^{\dagger }$ must be diagonal
 (possibly after removing null rows and rearanging the basis states), the matrices $\A$ and $\B$ are defined by the conditions
 that they diagonalize $\V \V ^{\dagger }$ and $\V ^{\dagger }\V $, respectively. 
$\V^\dagger \V$ has $\Nb$ generally nonzero eigenvalues $\EV_n^2$ ($n=1,2,\ldots ,\Nb$). 
The $\Na$-dimensional matrix $\V\V^\dagger$ has the same $\Nb$ eigenvalues as $\V^\dagger\V$ and additional $\Nd=\Na-\Nb$ zero eigenvalues.
From the vector form (\ref{V}) of $\V $ we obtain 
\begin{subequations}
\begin{eqnarray}
\mathsf{VV}^{\dagger } &=&\sum_{n=1}^{\Nb}\left\vert V_{n}\right\rangle
\left\langle V_{n}\right\vert ,  \label{VV+} \\
\V ^{\dagger }\V  &=&\left[ 
\begin{array}{cccc}
\left\langle V_{1}|V_{1}\right\rangle & \left\langle V_{1}|V_{2}\right\rangle
& \cdots & \left\langle V_{1}|V_{\Nb}\right\rangle \\ 
\left\langle V_{2}|V_{1}\right\rangle & \left\langle V_{2}|V_{2}\right\rangle
& \cdots & \left\langle V_{2}|V_{\Nb}\right\rangle \\ 
\vdots & \vdots & \ddots & \vdots \\ 
\left\langle V_{\Nb}|V_{1}\right\rangle & \left\langle
V_{\Nb}|V_{2}\right\rangle & \cdots & \left\langle
V_{\Nb}|V_{\Nb}\right\rangle
\end{array}
\right] .  \label{V+V}
\end{eqnarray}
\end{subequations}
$\V ^{\dagger }\V $ is the Gram matrix \cite{Gantmacher} for the set of vectors  $|V_{n}\rangle$, ($n=1,2,\ldots,\Nb$). 
Thus if all these vectors are linearly independent then $\det\V^\dagger\V \neq 0$ and all eigenvalues of $\V^\dagger\V$ are nonzero \cite{Gantmacher};
 however, this assumption of independence is unnecessary.

The MS Hamiltonian (\ref{H-MS}) has the explicit form
\begin{equation}
\widetilde{\H}(t) =\left[\begin{array}{c|c}
\mathsf{O} & \mathsf{O} \\ \hline
\mathsf{O} & 
\begin{array}{cccccccc}
0 & 0 & \cdots & 0 & \EV_1 f(t) & 0 & \cdots & 0 \\ 
0 & 0 & \cdots & 0 & 0 & \EV_2 f(t) & \cdots & 0 \\ 
\vdots & \vdots & \ddots & \vdots & \vdots & \vdots & \ddots & \vdots \\ 
0 & 0 & \cdots & 0 & 0 & 0 & \cdots & \EV_\Nb f(t) \\ 
\EV_1 f(t) & 0 & \cdots & 0 & \Delta & 0 & \cdots & 0 \\ 
0 & \EV_2 f(t) & \cdots & 0 & 0 & \Delta & \cdots & 0 \\ 
\vdots & \vdots & \ddots & \vdots & \vdots & \vdots & \ddots & \vdots \\ 
0 & 0 & \cdots & \EV_\Nb f(t) & 0 & 0 & \cdots & \Delta%
\end{array}
\end{array}
\right].  \label{H-MS explicit}
\end{equation}
The structure of $\widetilde{\H}(t)$ shows that in the MS basis the dynamics decomposes into sets of $\Nd$ decoupled single states
 and $\Nb$ independent two-state systems $|\MSa_n\rangle \leftrightarrow |\MSb_n\rangle $ ($n=1,2,\ldots ,\Nb$),
 each composed of a lower state $|\MSa_n\rangle $ and an upper state $|\MSb_n\rangle $, and driven by the two-state RWA Hamiltonians
\begin{equation}
\widetilde{\mathsf{h}}_n =\left[ \begin{array}{cc}
0 & \EV_n f(t) \\ 
\EV_n f(t) & \Delta%
\end{array}\right] , \quad (n=1,2,\ldots ,\Nb).  \label{Hn}
\end{equation}
Each of these two-state Hamiltonians has the same detuning $\Delta$, but they differ in the couplings $\EV _{n}$. 
Each of the new lower states $|\MSa_n\rangle $ is an eigenstate of $\V\V^\dagger$ that corresponds to a specific eigenvalue $\EV_n^2$ of $\V\V^\dagger$.
Similarly the new upper state $|\MSb_n\rangle $ is the eigenstate of $\V^\dagger\V$, corresponding to the same eigenvalue $\EV_n^2$. 
The square root of this common eigenvalue, $\EV_{n} $, represents the coupling (half the Rabi frequency) in the respective
 independent MS two-state system $|\MSa_n\rangle \leftrightarrow |\MSb_n\rangle $. 
The $\Nd$ zero eigenvalues of $\V \V ^{\dagger }$ correspond to decoupled (dark) states in the lower set
 (we assume thoughout that $\Na\geqq \Nb$; therefore, dark states, if any, are in the lower set). 
The dark states are decoupled from the dynamical evolution because they are driven by one-dimensional null Hamiltonians.


\subsection{The  propagator in the original basis}


The eigenvectors of $\V \V ^{\dagger }$ and $\V ^{\dagger }\V $ form the transformation matrices $\A$ and $\B$,
\begin{equation}
\A=\left[ \begin{array}{c}
\langle \MSd_1| \\ 
\vdots \\ 
\langle \MSd_{\Nd}| \\ 
\langle \MSa_1| \\ 
\vdots \\ 
\langle \MSa_{\Nb}|%
\end{array}%
\right] ,\quad \B=\left[ 
\begin{array}{c}
\langle \MSb_1| \\ 
\vdots \\ 
\langle \MSb_{\Nb}|%
\end{array}\right] .  \label{A B}
\end{equation}
They obey the completeness relations
\begin{subequations}\label{completeness}
\begin{eqnarray}
\sum_{n=1}^{\Nb}|\MSa_n\rangle \langle \MSa_n| +\sum_{k=1}^{\Nd}|\MSd_k\rangle \langle \MSd_k| = \I, \label{completeness a}\\ 
\sum_{n=1}^\Nb|\MSb_n\rangle \langle \MSb_n| =\I.  \label{completeness b}
\end{eqnarray}
\end{subequations}

The propagator $\widetilde{\u}_{n}(t_{f},t_{i})$ for the independent MS two-state system $|\MSa_n\rangle \leftrightarrow |\MSb_n\rangle $ ($n=1,2,\ldots ,\Nb$)
 is unitary and therefore can be parameterized in terms of the  complex-valued Cayley-Klein parameters $\CKa _{n}$ and $\CKb _{n}$, 
\begin{equation}
\widetilde{\u}_{n}(t_{f},t_{i})= \begin{bmatrix}
\CKa _{n} &  \CKb _{n} \\
-\CKb _{n}^{\ast }e^{-i\delta} & \CKa _{n}^{\ast }e^{-i\delta}\end{bmatrix},  \label{Un}
\end{equation}
where  $\left\vert \CKa_n\right\vert^2+\left\vert \CKb_n\right\vert^2=1$ and $\delta =\int_{t_i}^{t_f}\Delta (t^\prime)dt^\prime$. 
In what follows we express the desired control of statevector motion in terms of constraints on these two-state Cayley-Klein parameters.
The unimportant phase factor $e^{-i\delta}$ originates from the chosen representation of the Hamiltonian (\ref{H}),
 which facilitates the application of the MS transformation.
In the interaction representation, where the diagonal elements are nullified
 and the detunings appear in phase factors multiplying the couplings, the factor $e^{-i\delta}$ disappears.
(Note: This phase factor has been omitted in Eqs. (17), (18) and (20) of Ref. \cite{Kyoseva},
 where it should have been associated with $\CKa^\ast$ and $\CKb^\ast$; 
 however, it did not affect any result there.)

By taking into account the MS propagators (\ref{Un}) for the two-state MS systems, the ordering of the states (\ref{A B}),
 and the MS Hamiltonian (\ref{H-MS explicit}), we write the propagator of the full system in the MS basis as 
\begin{equation}
\widetilde{\U}= \left[\begin{array}{c|c}
\I & \mathsf{O} \\ \hline
\mathsf{O} &  
\begin{array}{cccccccc}
\CKa _{1} & 0 & \cdots & 0 & \CKb_{1} & 0 & \cdots & 0 \\ 
0 & \CKa _{2} & \cdots & 0 & 0 & \CKb_{2} & \cdots & 0 \\ 
\vdots & \vdots & \ddots & \vdots & \vdots & \vdots & \ddots & \vdots \\ 
0 & 0 & \cdots & \CKa _{\Nb} & 0 & 0 & \cdots & \CKb_{\Nb}\\ 
-\CKb _{1}^{\ast }e^{-i\delta} & 0 & \cdots & 0 & \CKa _{1}^{\ast }e^{-i\delta} & 0 & \cdots & 0 \\ 
0 & -\CKb _{2}^{\ast }e^{-i\delta} & \cdots & 0 & 0 & \CKa _{2}^{\ast }e^{-i\delta} & \cdots & 0 \\ 
\vdots & \vdots & \ddots & \vdots & \vdots & \vdots & \ddots & \vdots \\ 
0 & 0 & \cdots & -\CKb _{\Nb}^{\ast }e^{-i\delta} & 0 & 0 & \cdots & \CKa _{\Nb}^{\ast }e^{-i\delta}
\end{array}\end{array}\right] .  \label{U-MS}
\end{equation}

It is straightforward to show that the propagator in the original basis $\U=\mathsf{S}^{\dagger }\widetilde{\U}\mathsf{S}$ reads 
\begin{subequations}\label{U full}
\begin{equation}
\U=\left[ \begin{array}{cc}
\U_\Na & \U_{\Na\Nb} \\ 
\U_{\Nb\Na} & \U_\Nb%
\end{array}\right] ,  \label{U}
\end{equation}
where 
\begin{eqnarray}
\U_\Na &=&\sum_{n=1}^{\Nb}\CKa _{n}|\MSa_n\rangle \langle \MSa_n|+\sum_{k=1}^{\Nd}|\MSd_k\rangle \langle\MSd_k |,  \label{Uaa} \\
\U_{\Na\Nb} &=& \sum_{n=1}^{\Nb}\CKb_{n}|\MSa_n\rangle \langle \MSb_n|,  \label{Uab} \\
\U_{\Nb\Na} &=& -e^{-i\delta}\sum_{n=1}^{\Nb}\CKb_{n}^\ast|\MSb_n\rangle \langle \MSa_n|,  \label{Uba} \\
\U_\Nb &=& e^{-i\delta} \sum_{n=1}^{\Nb}\CKa _{n}^{\ast }|\MSb_n\rangle \langle \MSb_n|.  \label{Ubb}
\end{eqnarray}
\end{subequations}
The propagator $\U_\Na$ connects states within the lower set, $\U_\Nb$
connects states  within the upper set, and $\U_{\Na\Nb}$ and $\U_{\Nb\Na}$ mix states from the lower and upper sets. 
By using the completeness relations (\ref{completeness a}) and (\ref{completeness b}) we find 
\begin{subequations}
\begin{eqnarray}
 \U_\Na &=& \I +\sum_{n=1}^{\Nb}(\CKa_{n}-1)|\MSa_n\rangle \langle \MSa_n|,\\
 \U_\Nb &=& e^{-i\delta}\left[\I +\sum_{n=1}^{\Nb}(\CKa_{n}^\ast-1)|\MSb_n\rangle \langle \MSb_n|\right].
\end{eqnarray}
\end{subequations}
Hence the propagator $\U_\Na$ does not depend on the decoupled states $|\MSd_k\rangle $ $(k=1,2,\ldots ,\Nd)$. 
This has to be expected because, owing to their degeneracy, the choice of the decoupled states is not unique:
 any superposition of them is also a zero-eigenvalue eigenstate of $\V \V ^{\dagger }$.
Because the dynamics in the original basis must not depend on such a leeway of choice,
 the  full propagator $\U$ must not depend on the decoupled states at all, and this is indeed the case.


\subsection{Analytical solutions for degenerate levels}


Equation (\ref{U}) expresses the dynamics of the degenerate two-level system
 in terms of the dynamics of the $\Nb$ independent non-degenerate two-state systems, 
 each with the same detuning $\Delta (t)$ and pulse shape $f(t)$ of the couplings but with different coupling strengths  $\EV_n$. 
Therefore, Eq. (\ref{U}) allows one to generalize any analytical non-degenerate two-state solution to a pair of degenerate levels. 
Recently, such a generalization of the Landau-Zener model to two degenerate levels has been presented \cite{Vasilev}. 
This generalization displays several interesting properties, for instance, not all transition probabilities
 between degenerate states are defined for infinite time duration. 
Other analytical solutions involving two degenerate levels have been derived for five chainwise-coupled states
 in $M$ or $W$ linkage configurations \cite{Vitanov MW}. 
We present below another interesting aspect of the solution (\ref{U full}): its geometrical nature.


\section{Quantum-state reflections}



\subsection{Coupled reflections}


Of particular significance is the special case when the Cayley-Klein parameters $\CKb _{n}$ are all equal to zero,
\begin{equation}
\CKb _{n}=0\quad (n=1,2,\ldots ,\Nb);  \label{beta=0}
\end{equation}%
then all transition probabilities in the MS basis, as well as these in the original basis from the lower set to the upper set, vanish,
 i.e. $\U_{\Na\Nb}=\U_{\Nb\Na}=0$. 
Equation (\ref{beta=0}) implies that $\left\vert \CKa_{n}\right\vert =1$, or 
\begin{equation}
\CKa _{n}=e^{i\phi _{n}}\quad (n=1,2,\ldots ,\Nb),  \label{alpha_n}
\end{equation}%
for all MS two-state systems, where $\phi _{n}$ are arbitrary phases.

After substituting Eqs. (\ref{beta=0}) and (\ref{alpha_n}) in Eqs. (\ref{U full}), the propagator in the original basis reads
\begin{subequations}\label{U-QHR}
\begin{eqnarray}
\U&=& \left[ \begin{array}{cc}
\U_\Na & \mathsf{O} \\ 
\mathsf{O} & \U_\Nb
\end{array}\right] ,  \\
\U_\Na &=&\I +\sum_{n=1}^{\Nb}(e^{i\phi_{n}}-1)|\MSa_n\rangle \langle \MSa_n|,\\  \label{Ua}
\U_\Nb &=&e^{-i\delta} \left[ \I +\sum_{n=1}^{\Nb}(e^{-i\phi_{n}}-1)|\MSb_n\rangle \langle \MSb_n| \right],  \label{Ub}
\end{eqnarray}
\end{subequations}
where  the propagator $\U_\Na$ operates in the lower set of states and $\U_\Nb$ acts in the upper set.
Taking into account that the coupled lower states $|\MSa_n\rangle\ (n=1,2,\ldots,\Nb)$ are orthonormal basis vectors,
 i.e. $\langle \MSa_n|\MSa_k\rangle =\delta _{nk}$,
 as are the upper MS states $|\MSb_n\rangle$ $(n=1,2,\ldots,\Nb)$, we rewrite $\U_\Na$ and $\U_\Nb$ as the products 
\begin{subequations}
\begin{eqnarray}
\U_\Na &=&\prod_{n=1}^{\Nb}\M(\MSa_n,\phi _{n}),\label{Ua-M} \\
\U_\Nb &=&e^{-i\delta} \prod_{n=1}^{\Nb}\M(\MSb_n,-\phi_{n}),  \label{Ub-M}
\end{eqnarray}
\end{subequations}
where 
\begin{equation}
\M(\nu;\phi )=\I +(e^{i\phi }-1)|\nu\rangle \langle \nu|.
\label{Householder}
\end{equation}%
These individual matrices are, in fact, generalizations of the matrices used to produce Householder reflection in matrix computations \cite{Householder}. 
We refer to them as \emph{generalized quantum Householder reflection} (QHR) operators  \cite{Kyoseva,Ivanov,Ivanov2}.
The generalized QHR operator (\ref{Householder}) is unitary, $\M(\nu;\phi )^{-1}=\M(\nu;\phi)^\dagger =\M(\nu;-\phi)$,
 and its determinant has unit magnitude, $\det \M(\nu;\phi)=e^{i\phi}$.
For $\phi =0$ the QHR operator reduces to the identity, $\M(\nu; 0)=\I$,
 while for $\phi =\pi $, the QHR operator reduces to the standard reflection, 
\begin{equation}
\M(\nu;\pi)=\I-2|\nu\rangle \langle \nu|.  \label{stanQHR}
\end{equation}

The orthogonality of the QHR vectors $|\MSa_n\rangle $ in Eq. (\ref{Ua-M}) automatically ensures the commutation of the QHRs,
\begin{equation}
\left[ \M(\MSa_n,\phi _{n}),\M(\MSa_m,\phi_m)\right] =0.
\end{equation}%
Therefore their ordering in the product in Eq. (\ref{Ua-M}) is unimportant.
The same argument applies to Eq. (\ref{Ub-M}).

The importance of QHR derives from the fact that any $N$-dimensional unitary matrix can be decomposed into
 a set of at most $N$ generalized QHRs \cite{Ivanov}. 
The resulting ease with which pulse sequences can be designed to realize the QHR therefore enables one to synthesize
 any desired unitary transformation of a qunit state, for example a quantum Fourier transform \cite{Ivanov}
 or transition between any two pure or mixed qunit states \cite{Ivanov2}.

\subsection{Special case: orthogonal interaction vectors}

In the special case when the interaction vectors $\vert V_n\rangle$ are \emph{orthogonal},
 $\langle V_m|V_n\rangle =\vert V_n\vert^2\delta_{mn}$, the MS eigenvalues and the MS states simplify greatly. 
Then the matrix $\V^\dagger\V$ of Eq. (\ref{V+V}) becomes diagonal,
 and hence its eigenvalues are $\EV_n^2=\left\vert V_n\right\vert ^{2}$ ($n=1,2,\ldots ,\Nb$). 
Moreover, the eigenstates of $\V^\dagger\V$ -- the MS states in the upper set -- coincide with the original states,
\begin{equation}
|\MSb_n\rangle \equiv |\psib_{n}\rangle ,\quad (n=1,2,\ldots,\Nb).
\end{equation}
The coupled MS states in the lower set -- the eigenstates of $\mathsf{VV}^{\dagger }$ -- are readily found from Eq. (\ref{VV+}), 
\begin{equation}
|\MSa_n\rangle =\frac{1}{\left\vert V_{n}\right\vert }\left\vert V_{n}\right\rangle \equiv | \hat{V}_{n} \rangle .
\label{MS states orthogonal}
\end{equation}
The propagator $U_\Na$ in the lower set is a product of QHRs, with the normalized interaction vectors $|\hat{V}_n\rangle$ serving as QHR vectors,
 whereas the propagator $\U_\Nb$ in the upper set is a phase gate, 
\begin{subequations}
\begin{eqnarray}
\U_\Na &=& \prod_{n=1}^{\Nb}\M(\hat{V}_{n},\phi _{n}),\label{Ua ortho}\\
\U_\Nb &=& e^{-i\delta}\sum_{n=1}^{\Nb}e^{-i\phi _{n}}|\psib_n\rangle \langle \psib_n|. \label{Ub ortho}
\end{eqnarray}
\end{subequations}

The advantage of having orthogonal interaction vectors $\vert V_{n}\rangle $ is that they serve as QHR vectors. 
The implication is that in order to construct a pre-selected coupled-QHR transformation (\ref{Ua ortho}),
 the required couplings are directly obtained from Eq. (\ref{MS states orthogonal}). 
Otherwise, in the general case of non-orthogonal interaction vectors, a set of pre-selected QHR vectors $|\MSa_n\rangle$ ($n=1,2,\ldots,\Nb$),
 defined as the eigenvectors of $\mathsf{VV}^\dagger$, demand numerical derivation of the required couplings from Eq. (\ref{VV+}).


\subsection{Realizations}


\subsubsection{Off-resonant hyperbolic-secant pulses}

The condition (\ref{alpha_n}) can be realized with the Rosen-Zener model \cite{RZ,Kyoseva},
 which assumes constant detuning and hyperbolic-secant time dependence $f(t)$ for the couplings, with pulse duration $T$,
\begin{subequations}
\begin{eqnarray}
f(t) &=& \mbox{sech} (t/T),  \label{sech} \\
\Delta (t) &=& \text{const}.
\end{eqnarray}
The independent MS two-state systems share the same detuning $\Delta$ and pulse shape $f(t)$,
 but have different MS couplings $\EV_n$. 
For the Rosen-Zener model the Cayley-Klein parameters $\CKa_n$ ($n=1,2,...,\Nb$) read \cite{RZ,Kyoseva} 
\end{subequations}
\begin{equation}
\CKa _{n}=\frac{\Gamma^{2}\left( \frac12+\frac12i\Delta T\right)}
 {\Gamma \left( \frac12+\EV_{n}T+\frac12 i\Delta T\right) \Gamma \left(\frac12-\EV_{n}T+\frac12 i\Delta T\right) },  \label{alfa-rz}
\end{equation}%
where $\Gamma (z)$ is Euler's $\Gamma$-function. 
Using the reflection formula $\Gamma( 1/2+z) \Gamma(1/2-z) =\pi/\cos \pi z$, we find 
\begin{equation}
\left\vert \CKa _{n}\right\vert ^{2} = 1-\frac{\sin ^{2} (\pi\EV_{n}T) }{\cosh^{2}\left( \frac12\pi \Delta T\right) }.
\end{equation}%
It follows that the condition $\left\vert \CKa_{n}\right\vert =1$ is satisfied when $\EV _{n}T=l$ $( l=0,1,2,\ldots) $. 
The phase $\phi _{n}$ of $\CKa _{n}=e^{i\phi _{n}}$ depends on the detuning $\Delta $, but not on the corresponding coupling $\EV_n$,
 and for an arbitrary integer $l$ we find \cite{Kyoseva}
\begin{equation}
\phi_n = 2\arg \prod_{k=0}^{l-1}\left[ \Delta T+i\left( 2k+1\right) \right].
\end{equation}
Hence the QHR phase $\phi _{n}$ can be produced by an appropriate choice of the detuning $\Delta$.
This result shows that even though the couplings for the MS two-state systems are not the same,
 the phases $\phi_n$ of the Cayley-Klein parameters $\CKa _{n}$ coincide, $\phi_n\equiv \phi$.
This feature is unique for the sech pulse.
For other non-resonant pulses, e.g. Gaussian \cite{Vasilev-Gaussian}, the phase $\phi_n$ would depend also
 on the coupling and therefore will be generally different for each MS two-state system. 

\subsubsection{Resonant pulses}

For exact resonance ($\Delta=0$), the Cayley-Klein parameter $\CKa_n$ reads $\CKa_n=\cos(A_n/2)$,
 where the pulse area is $A_n = 2\EV_n \int_{-\infty}^{\infty}f(t)dt$.
When the pulse area is $A_n=2(2l+1)\pi $ ($l=0,1,2,\ldots$),
 the phase $\phi _{n}$ is equal to $\pi$; hence we obtain a physical realization for the standard QHR (\ref{stanQHR}). 
This result is not resticted to the sech pulse (\ref{sech}) but it is valid for any pulse shape with such an area \cite{Kyoseva,Ivanov}. 
When the pulse areas are multiples of $4\pi$, the phases $\phi_n$ vanish, $\CKa_n=1$, and the corresponding QHRs reduce to the identity.
Resonant pulses therefore do not produce variable QHR phases $\phi_n$, which can be used as free parameters.

\subsubsection{Far-off-resonant pulses}

Far-off-resonant pulses provide the opportunity for easy adjustment of the phase $\phi _{n}$, albeit only approximately.
Then the condition (\ref{alpha_n}) is fulfilled automatically
 because of the smallness of the transition probabilities in each MS system.
Specifically, if the common detuning $\Delta $ exceeds sufficiently much the largest MS coupling $\EV_{n}$,
 then all transition probabilities will be negligibly small, $\left\vert \CKb_{n}\right\vert \ll 1$. 
By adiabatic elimination of each upper MS state one finds 
\begin{equation}
\phi _{n} \approx\frac{\EV _{n}^{2}}{\Delta }\int_{-\infty }^{\infty}f^{2}(t)dt.  \label{phi n off}
\end{equation}
For a sech pulse the integral is equal to $2T$, and for a Gaussian to $\sqrt{\pi/2}$. 
Because each MS coupling $\EV _{n}$ is generally different, each phase $\phi _{n}$ will also be different. 
Any desired phase $\phi_n$, or a set of such phases, can easily be produced by choosing the original couplings $V_{mn}$,
 and hence the MS couplings $\EV_n$, appropriately.

\section{Two degenerate upper states}

Above, we described the dynamics of the degenerate two-level quantum system in the general case
 when the lower and upper levels had arbitrary degeneracies, $\Na$ and $\Nb$ respectively. 
In this section, we will illustrate these results with a specific example: when the upper set consists of just two degenerate states, i.e. $\Nb=2$. 
This case is insteresting because of the possible implementations in different real physical systems,
 several examples of which will be presented below.
Moreover, this special case allows for an elegant analytical treatment.


\subsection{General case}


We retain the notation for the lower states $\vert \psia_{n}\rangle $ $(n=1,2,...,N)$, and we denote the two
upper states $\vert \psib^{\prime }\rangle $ and $\vert\psib ^{\prime \prime}\rangle $. 
The interaction matrix (\ref{V}) reads 
\begin{equation}
\V =\left[ 
\begin{array}{cc}
V_{1}^{\prime } & V_{1}^{\prime \prime } \\ 
V_{2}^{\prime } & V_{2}^{\prime \prime } \\ 
\vdots & \vdots \\ 
V_{N}^{\prime } & V_{N}^{\prime \prime }%
\end{array}\right]
 \equiv \left[ \left\vert V^{\prime }\right\rangle ,\left\vert V^{\prime \prime }\right\rangle \right] ,  \label{V2}
\end{equation}
where $\left\vert V^{\prime }\right\rangle $ and $\left\vert V^{\prime\prime }\right\rangle $ are $N$-dimensional  
 interaction vectors comprising the couplings between the lower states and the corresponding upper state. 
The product $\V ^{\dag }\V $ reads%
\begin{equation}
\V^\dag\V =\left[\begin{array}{cc}
\left\vert V^{\prime }\right\vert^2 & \left\langle V^\prime|V^{\prime\prime}\right\rangle \\ 
\left\langle V^{\prime\prime}|V^\prime\right\rangle & \left\vert
V^{\prime\prime}\right\vert^2
\end{array}\right] .
\end{equation}

With the introduction of parameters $\theta$ and $\sigma$ through the definitions 
\begin{subequations}
\begin{eqnarray}
\frac{2\vert \langle V^{\prime }|V^{\prime \prime }\rangle\vert }
{\vert V^{\prime \prime }\vert ^{2}-\vert V^{\prime }\vert ^{2}} &=&\tan 2\theta \quad (0<\theta <\pi /2), \\
\arg \langle V^{\prime }|V^{\prime \prime }\rangle &=&\sigma ,
\end{eqnarray}
\end{subequations}
 we write the eigenvalues $\EV_m$ and the associated eigenvectors $|\MSb_m\rangle$ ($m=1,2$) within the upper set as
\begin{subequations}
\begin{eqnarray}
\EV_{1,2}^2 &=& \frac{\vert V^\prime\vert^2+\vert V^{\prime\prime}\vert^2}{2}
 \pm \frac{\vert V^{\prime}\vert^2 - \vert V^{\prime\prime}\vert^2}{2\cos 2\theta} ,  \label{eigenvalues}\\
|\MSb_1\rangle &=& \left[\begin{array}{c}\cos \theta \\ -e^{-i\sigma }\sin \theta\end{array}\right] ,\qquad
|\MSb_2\rangle = \left[ \begin{array}{c} e^{i\sigma }\sin \theta \\ \cos \theta\end{array}\right] .  \label{psi_b}
\end{eqnarray}
The next step is to find the MS states $|\MSa_1\rangle $ and $|\MSa_2\rangle $ within the lower set of states.
They are the eigenstates of the $\Na$-dimensional matrix, 
\end{subequations}
\begin{equation}
\V \V ^{\dag} = \vert V^{\prime}\rangle\langle V^{\prime}\vert + \vert V^{\prime\prime}\rangle \langle V^{\prime\prime}\vert ,  \label{vv+}
\end{equation}
which correspond to the (nonzero) eigenvalues (\ref{eigenvalues}). 
We construct them as superpositions of the interaction vectors $\vert V^\prime\rangle$ and $\vert V^{\prime\prime}\rangle$,
and find after simple algebra
\begin{subequations}
\label{psi_a}
\begin{eqnarray}
\vert\MSa_1\rangle &=& \frac{1}{\EV_1}\left(\cos\theta\left\vert V^\prime\right\rangle -e^{-i\sigma}\sin\theta\vert V^{\prime\prime}\rangle \right), \label{a-}\\
\vert\MSa_2\rangle &=& \frac{1}{\EV_2}\left(e^{i\sigma}\sin\theta\vert V^\prime\rangle +\cos\theta\vert V^{\prime\prime}\rangle \right). \label{a+} 
\end{eqnarray}
\end{subequations}

\subsection{The  propagators}

After we have found the explicit form of the MS states (\ref{psi_b}) and (\ref{psi_a}), we obtain the exact form of the propagators $\U_\Na$ and $\U_\Nb$, 
\begin{subequations}
\begin{eqnarray}
\U_\Na &=& \M(\MSa_1,\phi_1) \M(\MSa_2,\phi_2) \\
 &=& \I + (e^{i\phi_1}-1) |\MSa_1\rangle \langle\MSa_1\vert + (e^{i\phi_2}-1) \vert\MSa_2\rangle \langle \MSa_2\vert,\\
\U_\Nb &=& e^{-i\delta}\left[e^{-i\phi_1}\vert\MSb_1\rangle \langle \MSb_1\vert + e^{-i\phi_2}\vert\MSb_2\rangle \langle \MSb_2\vert \right],
\end{eqnarray}
\end{subequations}
with $\phi_1$ and $\phi_2$ being the phases of the Cayley-Klein parameters for the MS two-state propagators (\ref{Un}).

It is easy to verify that the bright states (\ref{a-}) and (\ref{a+}) are eigenstates of the propagator
 $\U_\Na$ with eigenvalues $e^{i\phi_1}$ and $e^{i\phi_2}$, respectively. 
Physically this means that if the qunit starts in one of these states it will end up in this same state, acquiring only a phase factor.
This occurs because of the conditions (\ref{beta=0}) and the independence of the different MS two-state systems. 
If the QHR phases are equal, $\phi_1=\phi_2$, then any superposition of $|\MSa_1\rangle$ and $|\MSa_2\rangle$ is also an eigenvector of $\U_\Na$. 
The other eigenstates of the propagator are all degenerate, with unit eigenvalue, and they are orthogonal to $|\MSa_1\rangle$ and $|\MSa_2\rangle$. 
For a qutrit ($\Na=3$) there is only one such eigenvector (up to an unimportant global phase factor)
 and it is proportional to $|\MSa_1\rangle \times|\MSa_2\rangle $. 
For higher-dimensional qunits, any vector in a hyperplane orthogonal to $|\MSa_1\rangle $ and $|\MSa_2\rangle $ is an eigenvector of $\U_\Na$.

In the special case when the vectors $\left\vert V^{\prime }\right\rangle $ and $\left\vert V^{\prime\prime}\right\rangle$ are orthogonal,
 $\left\langle V^{\prime }|V^{\prime \prime }\right\rangle =0$, the expressions  simplify. 
Then $\theta =0$ and $\sigma =0$, and hence 
\begin{subequations}
\begin{eqnarray}
\EV_1 &=& \vert V^{\prime}\vert ,\quad \EV_2 =\vert V^{\prime\prime}\vert , \\
\vert\MSa_1\rangle &=& \vert \hat{V}^{\prime}\rangle ,\quad \vert\MSa_2\rangle = \vert \hat{V}^{\prime\prime}\rangle , \\
\vert\MSb_1\rangle &=& \vert\psib_1\rangle,\quad \vert\MSb_2\rangle = \vert\psib_2\rangle.
\end{eqnarray}
\end{subequations}
Then the propagator $\U_\Na$ in the lower set is a product of QHRs, in which the interaction vectors
 $\ket{V^\prime}$ and $\ket{V^{\prime\prime}}$ serve as QHR vectors,
 while the propagator $\U_\Nb$ in the upper set is a phase gate,
\begin{subequations}
\begin{eqnarray}
\U_\Na &=& \M(\hat V^{\prime},\phi_1)\M(\hat V^{\prime\prime},\phi_2),\\
\U_\Nb &=&  e^{-i\delta} \begin{bmatrix} e^{-i\phi_1} & 0 \\ 0 & e^{-i\phi_2} \end{bmatrix}.
\end{eqnarray}
\end{subequations}

\subsection{Examples}

Following are examples of linkage patterns amongst angular-momentum states, which allow application of the QHR theory.


\subsubsection{Two levels, $J=3/2\leftrightarrow J=1/2$}


\begin{figure}[tb]
\includegraphics[width=70mm]{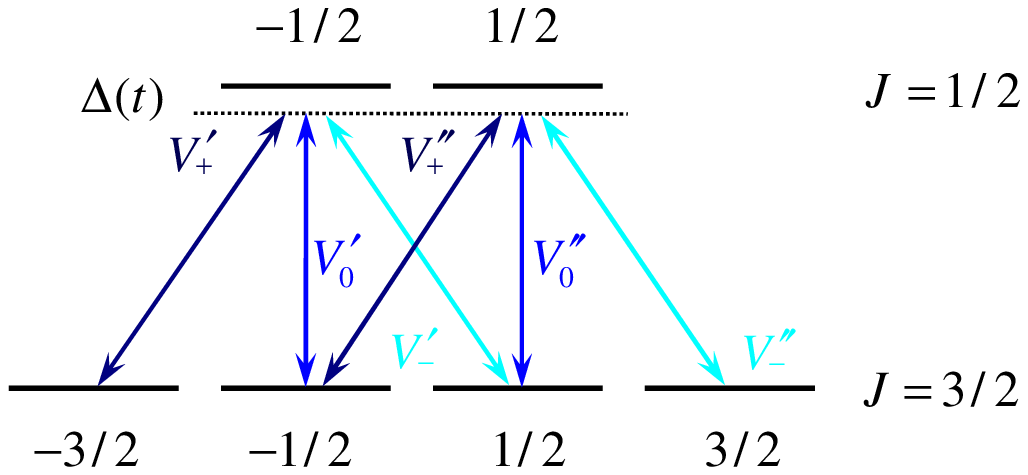}
\caption{(Color online) Linkage pattern for the four degenerate magnetic sublevels of $J=3/2$, the lower set,
 shown coupled by arbitrary polarization of electric-dipole radiation to the two sublevels of $J=1/2$, the upper set.  
The states are labeled by their magnetic quantum number.
}
\label{Fig-J3h1h}
\end{figure}

Figure \ref{Fig-J3h1h} shows linkage patterns possible with arbitrary polarization between the four magnetic sublevels of $J=3/2$, the lower set,
 and the two of $J=1/2$, the upper set.
The interaction matrix has the elements (with the Clebsch-Gordan coefficients included),
\begin{equation}
\V = \frac{1}{\sqrt{6}} \begin{bmatrix}
\sqrt{3} V_+ & 0 \\ 
-\sqrt{2} V_0 & V_+\\ 
V_- & -\sqrt{2}V_0\\ 
0 & \sqrt{3} V_-
\end{bmatrix},
\end{equation}
where the subscripts $+$, $-$ and 0 refer to right circular ($\sigma^+$), left circular ($\sigma^-$) and linear ($\pi$) polarizations.

\begin{figure}[tb]
\includegraphics[width=80mm]{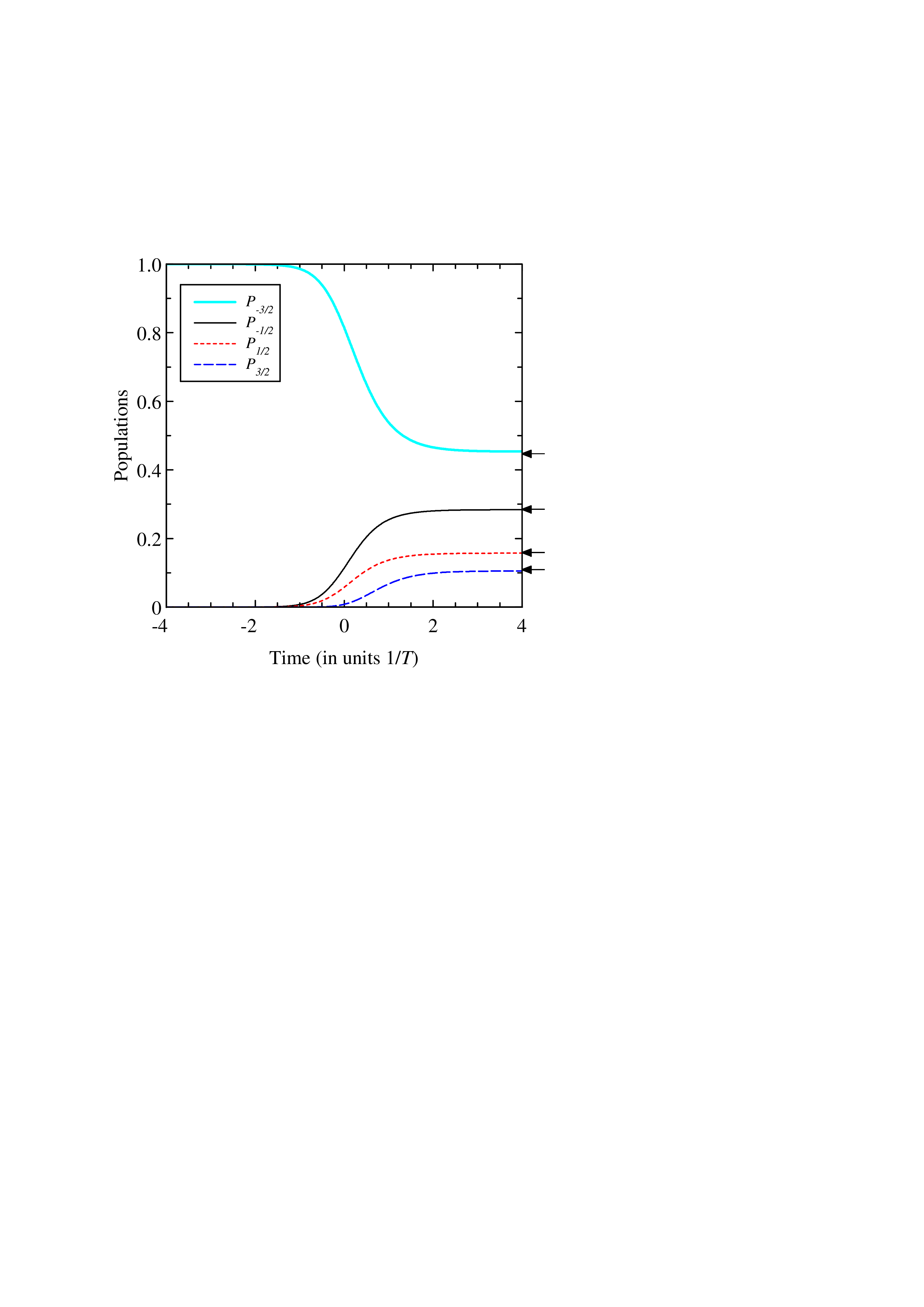}
\caption{(Color online) Time evolution of the numerically calculated populations of the magnetic sublevels of a $J=3/2$ level
 coherently coupled to a $J=1/2$ level by three polarized ($\sigma^+$, $\sigma^-$ and $\pi$) pulsed laser fields,
 with sech shape $f(t)=\text{sech}(t/T)$, $V_-=V_0=V_+=8.5T^{-1}$ and detuning $\Delta=80T^{-1}$.
Then $\theta=\pi/4$, $\sigma=\pi$, $\EV_1T=106.3$, $\EV_2T=38.2$, $\phi_1=2.65772$, $\phi_2=0.954776$. 
The arrows on the right indicate the values derived by the QHR theory.
}
\label{Fig-populations}
\end{figure}

Figure \ref{Fig-populations} shows an example of time evolution of the populations of the magnetic sublevels in the $J=3/2$ level,
 starting with all population in state $\ket{\psia_{-3/2}}$.
The conditions $\CKa_n=e^{i\phi_n}\ (n=1,2)$ are realized approximately, by using large detuning from the upper $J=1/2$ level.
In the end of the interaction, the numerically calculated populations are seen to approach the values predicted by the QHR theory (the arrows on the right).

\subsubsection{Two levels, $J=2\leftrightarrow J=1$}

\begin{figure}[tb]
\includegraphics[width=80mm]{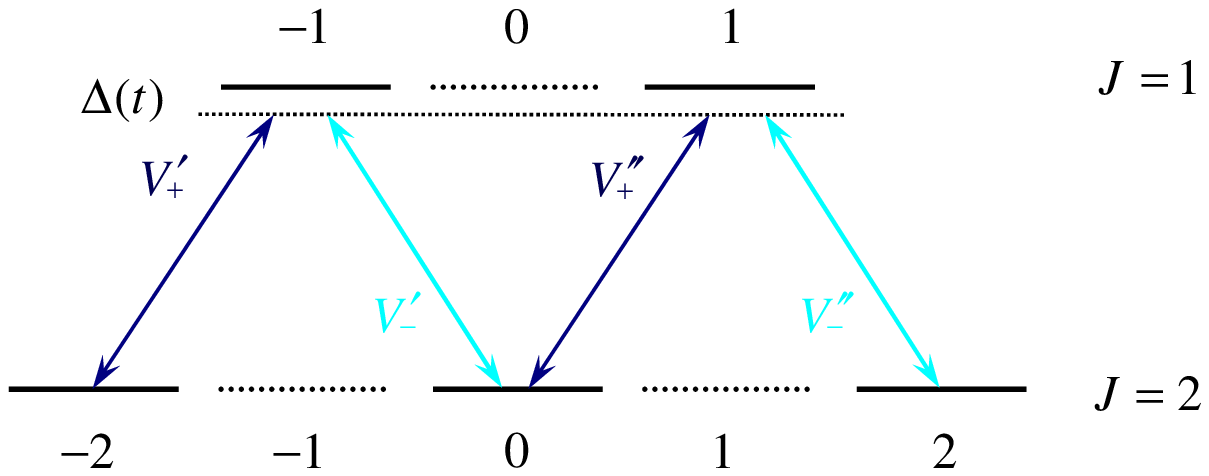}
\caption{(Color online) Linkage pattern for five states in an $M$ configuration. }
\label{Fig-J2J1}
\end{figure}

The linkage pattern for electric-dipole couplings between sublevels of $J=2$, the lower set, and $J=1$, the upper set will,
 for polarization expressed as a combination of left- and right-circular polarization, appear as two uncoupled systems:
 three states form a $\Lambda$-linkage, while five form an M-linkage, as shown in Fig. \ref{Fig-J2J1}. 
The interaction matrix is 
\begin{equation}
\V =\frac{1}{\sqrt{10}} \begin{bmatrix}
\sqrt{6}V_+ & 0 \\ 
V_- & V_+ \\ 
0 & \sqrt{6}V_-
\end{bmatrix}.
\end{equation}
Then $\tan 2\theta = |V_+V_-|/[5(|V_-|^2-|V_+|^2)]$ and $\sigma = \arg V_+ - \arg V_-$.

\subsubsection{Three levels, $J=0\leftrightarrow J=1\leftrightarrow J=0$}

\begin{figure}[tb]
\includegraphics[width=55mm]{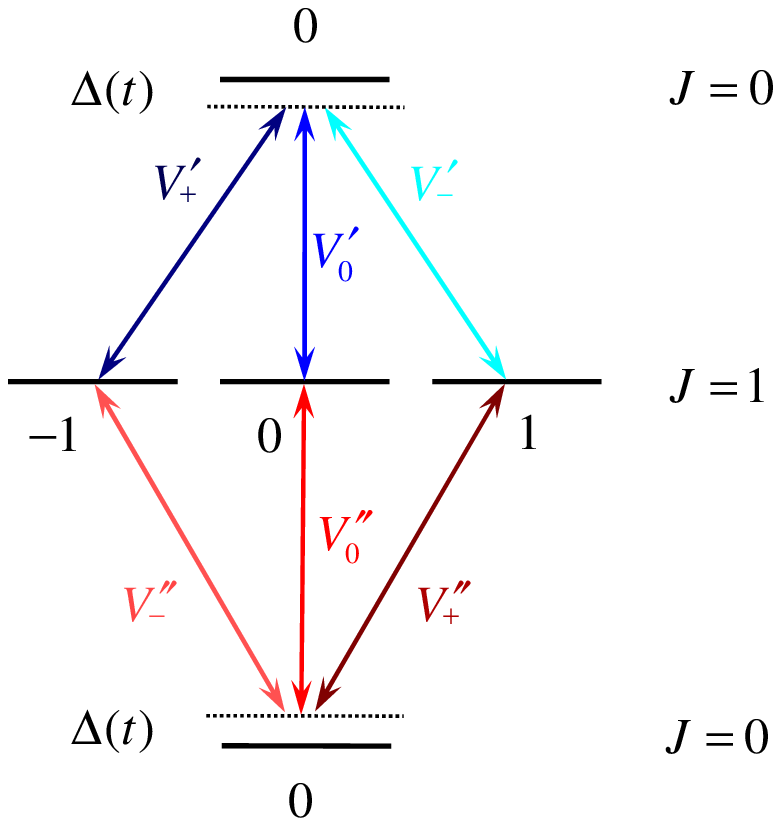}
\caption{(Color online) Linkage pattern for three-level ladder involving a degenerate middle level. 
The two ends of the chain, magnetic sublevels with $J=0$ have the same detuning from the three intermediate sublevels of $J=1$.
The various linkages are invoked by adjusting the direction of the polarization with respect to the quantization axis. 
The sublevels with $J=1$ form the lower set, while those with $J=0$ form the upper set.}
\label{Fig-J0J1J0}
\end{figure}

The MS transformation can be applied not only to a pair of degenerate levels but also to a ladder of degenerate levels,
 as long as there is only a single detuning. 
Figure \ref{Fig-J0J1J0} illustrates an example in an angular momentum basis between the magnetic sublevels of three levels with angular momenta $J=0,1,0$. 
The interaction matrix has the form 
\begin{equation}
\V =\begin{bmatrix}
V_{+}^{\prime } & V_{+}^{\prime \prime } \\ 
V_{0}^{\prime } & V_{0}^{\prime \prime } \\ 
V_{-}^{\prime } & V_{-}^{\prime \prime }.
\end{bmatrix},
\end{equation}
and the formalism of this section applies. 
In this case all fields can be changed independently.

\section{Discussion and conclusions}

We have here extended the earlier work on QHR in Hilbert space to allow more general linkage patterns between the quantum states,
 with particular attention to degenerate sublevels that occur with angular momentum states.
The extension relies on the use of the Morris-Shore transformation to reduce the original multi-linkage Hamiltonian
 to a set of independent two-state systems, thereby allowing the utilization of various known two-state analytic solutions.  
Such solutions, when expressed in terms of Cayley-Klein parameters, readily lead to conditions
 upon the pulse areas and the time-varying detunings of the excitation pulses.

We have found that the propagator within the lower (or upper) set of degenerate states represents coupled quantum-state mirrors.
The realization of these coupled QHRs within the lower (or upper) set of states requires certain conditions on the interaction parameters,
 specifically that all transition probabiities between the lower and upper sets must vanish.
We have proposed three physical realizations of this condition: with resonant, near-resonant (hyperbolic-secant), and far-off-resonant pulses.
Resonant and near-resonant pulses provide exactly zero transition probabilities, but offer less flexibility in the QHR phases $\phi_n$;
 moreover they require a carefully chosen pulse area in each MS two-state system, which leads to a number of conditions on the interaction parameters.
Far-off-resonant pulses fulfill the zero-probability conditions only approximately, but offer much more flexibility.
Then the only restriction is for sufficiently large detuning, without specific constraints on pulse areas,
 because the zero-probability conditions are fulfilled simultaneously in all MS two-state systems.

For angular-momentum states, there are six independent interaction parameters:
 three polarization amplitudes, two relative phases between different polarizations, and the common detuning.
Therefore the constructed QHR has six free parameters (with the far-off-resonance realization).
For a more general linkage, the number of independent parameters can be much larger.

In a subsequent paper \cite{multiQHR}, we shall describe the mathematical aspects of this largely unknown operator of coupled reflections,
 which however, as we have shown here, arises naturally in quantum systems.
In particular, we shall show how one can factorize, and therefore synthesize, an arbitrary U($N$) propagator by such objects.
The procedure is more efficient than a set of rotations that would produce the same transformation. 

We conclude by pointing out that the confinement of the statevector evolution to the lower set of states,
 and the availability of simple and efficient tools for its engineering, such as coupled QHRs,
 can be an essential ingredient for decoherence-free quantum computing \cite{decoherence}.

\acknowledgments 

Quantum information processing, and the needed manipulation of statevectors in Hilbert space, has long interested Sir Peter Knight,
 in whose honor the present issue of Journal of Modern Optics has been assembled. 
We are pleased to offer the present article in his honor.

This work has been supported by the EU ToK project CAMEL (Grant No.~MTKD-CT-2004-014427),
 the EU RTN project EMALI (Grant No. MRTN-CT-2006-035369), and the Alexander von Humboldt Foundation.


\end{document}